# Comment on "Half metallicity along the edge of zigzag boron nitride nanoribbons (Phys. Rev. B 78, 205415 (2008))"

Hari Mohan Rai[1*], Rajesh Kumar[1], Pankaj R. Sagdeo[1]

[1]*Material Research Lab. (MRL), Indian Institute of Technology Indore, Indore (M.P.) – 452020, India.*

Neeraj K. Jaiswal[2]

[2]*PDPM- Indian Institute of Information Technology, Design and Manufacturing, Jabalpur – 482005, India*

and Pankaj Srivastava [3]

[3]*Computational Nanoscience and Technology Lab. (CNTL), ABV- Indian Institute of Information Technology and Management, Gwalior – 474015, India*

We would like to comment that the prediction of 'Half-mtallicity' in only B edge H-passivated zigzag boron nitride nanoribbons (ZBNNR-BH), by Zheng et al.[1], is not correct as their interpretation is erroneous. Since it is well known that for a material, to be a half-metal, as a primary condition one type of spin channels or bands (either spin up or spin down) are conducting and the opposite ones are insulating. In commented article [1], ZBNNR-BH shows insulating behavior for spin up bands ($E_g$=4.5 eV) whereas conducting nature for spin down electrons. They claim[1] through the electronic band structure of 8-ZBNNR-BH [as shown in Figure 2 (b)][1], that spin down - α (conduction band minimum - CBM) and β (valence band maximum - VBM) bands are crossing each other at the Fermi level (green dotted line) and thus the ribbons are conducting for spin down electrons. But this 'crossing' of two same spin bands is impossible because according to well-known Pauli's exclusion principle- "Two identical fermions (particles with half-integer spin) cannot occupy the same quantum state simultaneously". In the case of electrons, "it is impossible for two electrons of a poly-electron atom to have the same values of the four quantum numbers ($n$, $\ell$, $m_\ell$ and $m_s$)". Thus for two electrons residing in the same orbital (same energy level e.g. above stated 'crossing' point [1]), $n$, $\ell$, and $m_\ell$ are the same, so $m_s$ must be different and the electrons must have opposite spins. Authors[1] have given an argument that - "the wave functions of α and β bands have different symmetries: being symmetric and antisymmetric with respect to the basal plane of the BN sheet" and the said crossing of α and β bands has been attributed to opposite symmetries (symmetric and antisymmetric) of wave functions corresponding to α and β bands[1]. But, we counter this because, as α and β both are fermionic (electronic) bands therefore presence of opposite symmetries for corresponding wave functions violates Pauli's exclusion principle. Let us assume two wave functions $\psi_\alpha$ and $\psi_\beta$ corresponding to two electrons (for said $\alpha$ and $\beta$ bands) present in ZBNNRs under study. Since all B and N atoms in ribbon are connected with each other, therefore, it is essential to analyze the exchange symmetry because, the commented crossing[1] is possible only when the state (energy state) corresponding to combined/overall wave function $\psi_{\alpha\beta}$ exist. The combination of $\psi_\alpha$ and $\psi_\beta$ i.e., two-particle wave function in terms of single-particle wave function is given by following equations (page No. 465 of ref. 2)

$$\Psi_{s(\alpha\beta)} = \frac{1}{\sqrt{2}}(\Psi_\alpha * \Psi_\beta + \Psi_\beta * \Psi_\alpha)$$

$$\Psi_{a(\alpha\beta)} = \frac{1}{\sqrt{2}}(\Psi_\alpha * \Psi_\beta - \Psi_\beta * \Psi_\alpha) \tag{1}$$

Further, in the case of fermions the space and spin parts must have different parities, leading to an overall wave function that is antisymmetric (page No. 468 of ref. 2) of the following form,:

$$\Psi_{a(\alpha\beta)} = \frac{1}{\sqrt{2}}[(\Psi_\alpha * \Psi_\beta - \Psi_\beta * \Psi_\alpha)]\chi_a(\vec{S}_\alpha, \vec{S}_\beta) \tag{2}$$

Here, $1/\sqrt{2}$ is for normalization. $\chi_a(\vec{S}_\alpha, \vec{S}_\beta)$ is the antisymmetric spin part which is a singlet (page No. 469 of ref. 2). For commented article[1], both $\psi_\alpha$ and $\psi_\beta$ are corresponding to same spin electrons, therefore, $\psi_\alpha = \psi_\beta$ and consequently

$$\Psi_\alpha * \Psi_\beta = \Psi_\beta * \Psi_\alpha \tag{3}$$

Now, it is obvious from above equations (2) and (3) that $\Psi_{a(\alpha\beta)} = 0$; means no antisymmetric wave function exists for considered case which implies that no such state (energy) is present in (at) which two same spin electrons (electronic bands) can exist simultaneously. Thus, the reason given by Zheng et al.[1] for the crossing of $\alpha$ and $\beta$ bands ("The wave functions of the $\alpha$ and $\beta$ bands have different symmetries: being symmetric and antisymmetric with respect to the basal plane of the BN sheet. These explain why the $\alpha$ and $\beta$ bands in Fig. 2 (b)[1] cross each other without having to create a band gap.") is not convincing and hence the commented crossing[1] is not possible though it is allowed between the bands with opposite spins. Our counter argument is also supported by J.M. Purneda[3]; he has also shown the same situation for C-BN heterostructures (Figure 3) and due to the presence of non-zero gap between same spin bands across the fermi level, he designated these structures as half semi-metals instead of Half-metals[3]. In order to further justify our comment we have reproduced the results by calculating electronic band structure and density of states (DOS) for same ZBNNR-BH configuraion[1] through DFT based Atomistix Tool Kit-Virtual NanoLab (ATK-VNL) [4–6]. The results for electronic band structure are found to be exactly identical with commented results[1].The difference between present and commented results is in their presentation and interpretation; therefore, we displayed electronic band structure (present Figure 1 and Figure 2) for spin down bands only as the band structures corresponding to spin up states[1] completely matches with present results whereas the DOS profile is slightly differ as our results show complete absence of any DOS peak at fermi level [6] which is also supported by transmission data [6]. A magnified view of above stated crossing[1] of spin down bands is depicted in the inset of present Figure 1 which clearly shows the existence of a non-zero band gap across the Fermi level and it is clearly visible that none of the band is crossing Fermi level as α and β bands are lying completely in CB and VB respectively. Moreover, present observation is not an artifact as we have checked this for different systems and not only across the fermi level, but also at different positions in E-K diagrams. Figure 2 shows the magnified views of various expected crossing points at various positions in E-K diagrams. Without magnifying the band structure plot, it appears that same spin electronic bands are also crossing each other but when crossing point is magnified then it is observed that none of the same spin electronic bands is crossing each other at any point as depicted in Figure 2. Moreover, the DOS profile, shown in Figure 2 (c), [1] also points towards a non-zero band gap for spin down electrons in the form of appearance of a small energy valley (window) across the Fermi level which is also supported by our DOS and transmission data[6]. Conclusively, we comment that by the definition the commented 8-ZBNNR-BH are ' semi-metals/half semi-metals [3,6] with non-zero band gap unlike to 'half-metals'[1], as primary requisite for half-metallicity i.e. proper crossing of fermi level by either only spin up or only spin down electronic band (s) [7-9] is completely absent[1].

## Acknowledgements

The authors are thankful to Computational Nanoscience and Technology Laboratory (CNTL), ABV—Indian Institute of Information Technology and Management Gwalior (India) for providing computational facility. One of the authors (HMR) acknowledges the Ministry of

Human Resource Development (MHRD), government of India for providing financial support as Teaching Assistantship.

## References

[1] F. Zheng, G. Zhou, Z. Liu, J. Wu, W. Duan, B.-L. Gu, and S.B. Zhang, Phys. Rev. B **78**, 205415 (2008).

[2] N. Zettili,Quantum Mechanics Concepts and Applications 2$^{nd}$ Ed., Chapter **8,.** (Wiley, New York, 2009).

[3] J.M. Purneda, , Phys. Rev. B **81**, 161409 (R) (2010).

[4] M. Brandbyge, J.-L. Mozos, P. Ordejón, J. Taylor, and K. Stokbro, Phys. Rev. B **65**, 165401 (2002).

[5] M. Chattopadhyaya, M.M. Alam, S. Sen, and S. Chakrabarti, Phys. Rev. Lett. **109**, 257204 (2012).

[6] H.M. Rai, N.K. Jaiswal, P. Srivastava, and R. Kurchania, J. Comput. Theor. Nanosci. **10**, 368 (2013).

[7]Y. Wang, Y. Ding, and J. Ni, Phys. Rev. B **81**, 193407 (2010).

[8] H.M. Rai, S.K. Saxena, V. Mishra, R. Late, R. Kumar, P.R. Sagdeo, N.K. Jaiswal, and P. Srivastava, Solid State Commun. **212**, 19 (2015).

[9] H.M. Rai, S.K. Saxena, V. Mishra, R. Late, R. Kumar, P.R. Sagdeo, N.K. Jaiswal, and P. Srivastava, RSC Adv. **6**, 11014 (2016)

## Figure Captions

Figure 1; (Color online) Calculated electronic band structure corresponding to spin down states for 8-ZBNNR-BH configuration.Inset shows the magnified view of commented crossing point.

Figure 3; (Color online) (a) and (b) are showing magnified view of different expected crossing points (CP). In (b) different CPs are enclosed under dotted rectangles with numbering 1 to 5. (c) – (g) show the magnified views of these CPs whereas, magnified views of CPs enclosed under dotted rectangles in (g) are depicted in (h) – (j). It is clear from these Figures that same spin electronic bands do not cross each other at any point.

Figure 3; (Color online) Calculated electronic band structure corresponding to spin up and spin down states for C-BN heterostructures. The crossing point for red bands is magnified and pointed out with an arrow. (**Phys. Rev. B 81, 161409 (R) (2010)**))

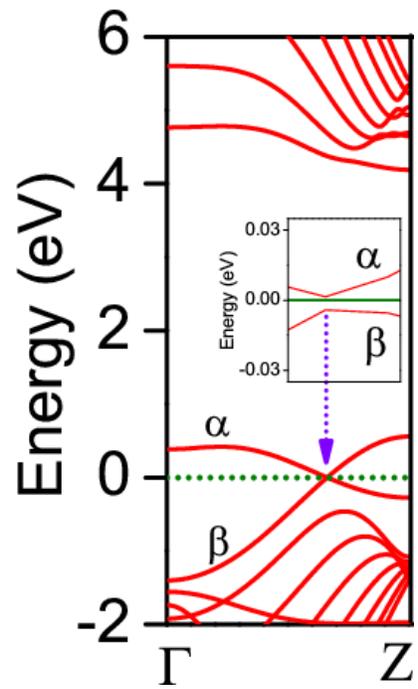

**Figure 1**    **Rai et al.**

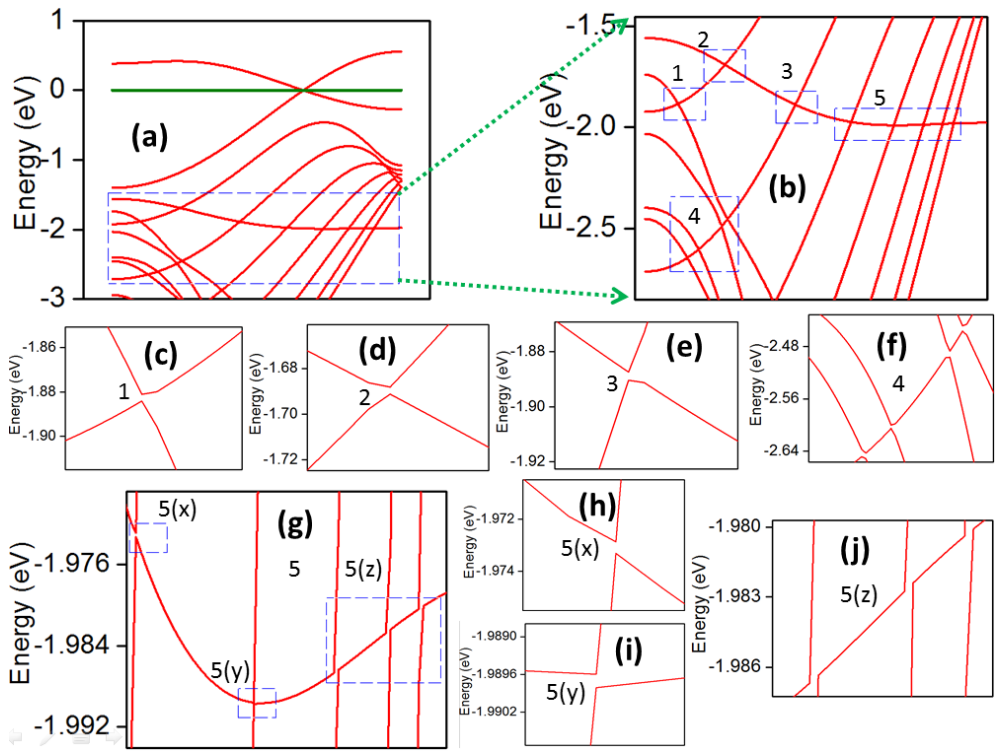

**Figure 2**    **Rai et al.**

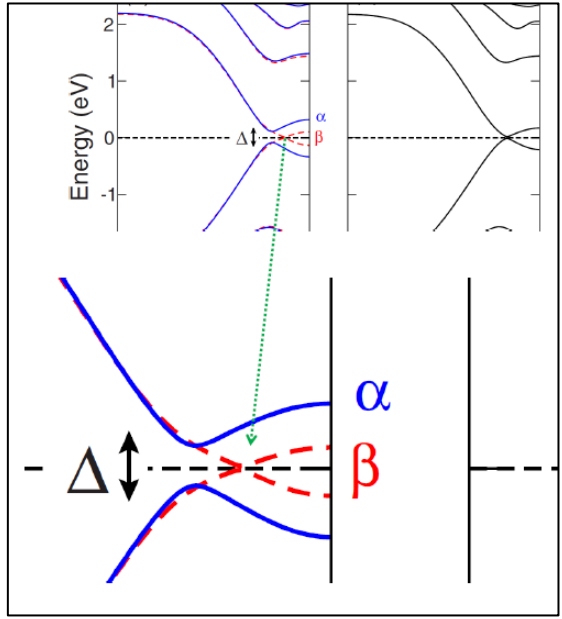

**Figure 3**    **Phys. Rev. B 81, 161409 (R) (2010)**